\documentclass[aps,floatfix,prd,showpacs,twocolumn]{revtex4}

\usepackage{graphicx}
\usepackage{dcolumn}
\usepackage{bm}

\begin{document}

\def\be{\begin{equation}}
\def\ee{\end{equation}}
\def\lr{\leftrightarrow}

\title{Confinement Models at Finite Temperature and Density}
\author{Pok Man Lo and Eric S. Swanson}
\affiliation{
Department of Physics and Astronomy, 
University of Pittsburgh, 
Pittsburgh, PA 15260, 
USA.}

\date{\today}

\begin{abstract}
In-medium chiral symmetry breaking in confining potential models of QCD is examined. 
Past attempts to analyse these models have been hampered by infrared divergences that
appear at non-zero temperature. 
We argue that previous attempts to circumvent this problem are not satisfactory and demonstrate
a simple resolution. We also show that the expectation that confining models do not exhibit a 
chiral phase transition  is incorrect.  The effect of summing ring diagrams is investigated and we present the first determination
of the temperature-density phase diagram for three model systems.
We find that observables and the phase structure of the confinement models depend 
strongly on whether vacuum polarisation is accounted for. Finally, it appears that standard  confinement 
models cannot adequately describe both hadron phenomenology and in-medium properties of 
QCD. 
\end{abstract}
\pacs{12.38.Lg, 12.38.Mh, 12.39.Pn, 21.65.Qr}

\maketitle

\section{Introduction}

The properties of QCD at finite temperature and density find applications in topics as diverse
as the nature of proto-neutron stars, early universe cosmology, and experiment at RHIC and the LHC\cite{AAA}.
Unfortunately, many of the properties of interest are nonperturbative, and, with the exception of lattice techniques,  tools for dealing with
nonperturbative field theory remain rudimentary. Heightening our discomfort is the continuing statistical `minus sign' problem in lattice field theory that is present at finite chemical potential\cite{BBB}. Furthermore, well-known problems of infrared (IR) divergences at high order in perturbation theory persist\cite{CCC}.
Even old hopes that QCD at large temperature is perturbative may be misplaced since large temperature QCD corresponds to the dimensionally reduced QCD of three dimensions, which is also confining. Thus it is of interest to develop nonperturbative methods for analysing QCD and model systems in medium. 

Here we examine the properties of two simple models of confinement that are motivated by QCD  in Coulomb gauge. 
The analysis of such models dates from the mid-1980s and parallels extensive work with Nambu-Jona-Lasinio (NJL)  models\cite{K}. 
(We note that NJL models are not renormalisable and do not exhibit confinement.)   In the following we will examine a contact model in detail, focussing on coupling constant dependence of
the phase diagram, possible critical points, and the effects of incorporating polarisation in the formalism. We demonstrate that a surprisingly rich structure emerges.

We also examine confining potential models with the intent of clarifying several points in the literature. The first of these concerns the existence of infrared divergences in the temperature-dependent gap equations that naively renders them useless. The second issue is whether a linear model can support a chiral phase transition.  We resolve these issues and also examine the effects of polarisation on the phase diagram. 
As far as we are aware this investigation presents the first determination of the phase structure of confining models and the first examination of general vacuum polarisation effects in  
contact and confining models.

\section{Confinement Models and Infrared Divergences}

\subsection{Linear and Contact Models}
\label{models}

Our starting point is the Hamiltonian of QCD in Coulomb gauge. Coulomb gauge is especially appropriate for the study of in-medium properties of QCD because all its degrees of freedom are physical. The
usual demerit associated with non-manifest covariance is obviated by the presence of the heat bath.
Upon neglecting transverse gluons, the QCD Hamiltonian takes the form

\begin{equation}
H = \int \bar\psi (- i \vec \gamma \cdot \nabla + m) \psi  +\frac{1}{2} \int \rho^a(x) V(x-y) \rho^a(y)
\end{equation}
where $\rho^a = \psi^\dagger T^a \psi$ is the colour quark current and $T^a$ is a generator of $SU(N)$.

Neglecting transverse gluons ruins the gauge and Lorentz invariance properties to the theory. However, it has been argued that (dynamically) massive transverse gluons provide a more accurate basis
for the exploration of low energy properties of QCD\cite{ss7,C}. Thus, as long as the induced error
is not too great, gauge invariance should be approximately respected. Similarly, the full theory is Lorentz invariant, even if it is not manifestly covariant. Accurate truncations should accurately  preserve this property.

The instantaneous interaction kernel corresponds to the Coulomb potential in QED; in the case of QCD it
can be written as the vacuum expectation value of the Coulomb operator\cite{tdl, ss7}

\begin{equation}
\delta^{ab}\, V(\vec x - \vec y) = \langle\Omega | (\vec x a| \frac{g}{\nabla\cdot D} (-\nabla^2)\frac{g}{\nabla\cdot D}| \vec y b) | \Omega \rangle.
\end{equation}
Here $\vec D^{ab} = \vec \nabla \delta^{ab} +  g f^{acb}\vec A^c$ is the adjoint covariant derivative, $a$, $b$ are colour indices, and $\Omega$ is the full vacuum. This interaction is often modelled as a linear confinement potential:

\begin{equation}
V(\vec r) = - \frac{3}{4} b r, \qquad \qquad V(\vec q) = \frac{6 \pi b}{q^4}
\label{LinEq}
\end{equation}
The string tension is denoted $b$ and its phenomenological value is  approximately $0.2$ GeV$^2$.

An alternative that matches to perturbation theory is provided by the Richardson potential, 
\begin{equation}
V(\vec q) = \frac{3}{4} \frac{4\pi}{q^2 \beta_0 {\rm log}(1 + q^2/\Lambda^2)}
\label{RichEq}
\end{equation}
with  $\beta_0 = 11-\frac{2}{3}n_f$, $\Lambda^2 = 2 b \beta_0$, and $n_f$ is the number of quark flavours.

Finally, we shall consider a simple contact model defined by

\begin{equation}
V(\vec r) = \frac{\lambda}{\Lambda^2} \delta(\vec r).
\end{equation}
The scale $\Lambda$ is introduced to set dimensions and will be used as an ultraviolet cutoff in this model. Of course the contact model is not confining, however it permits detailed analysis of in-medium effects since the resulting equations are considerably simplified. It also serves as a reference point for the confining models.

The partition function is defined as

\begin{equation}
Z[\bar \eta, \eta] = \int D\bar \psi D\psi \, {\rm exp}[- A + \bar \eta \psi + \bar \psi \eta]
\end{equation}
with
\begin{eqnarray}
A  &=& \int_0^\beta d\tau d^3 x \, \bar \psi( \gamma_0 (\partial_\tau - \mu) - i \vec \gamma \cdot \vec \nabla + m) \psi + \nonumber \\
&&\!\!\!\!\!\!\!\!\!\!\!\!\! 
\frac{1}{2} \int_0^\beta d\tau d^3x d\tau' d^3 y \, \rho^a(x) V(\vec x -\vec y) \delta(\tau-\tau') \rho^a(y)
\end{eqnarray}
Notice that a quark chemical potential term, proportional to $\mu$,  has been added to the action.

We employ the imaginary time formalism for evaluating the partition function. In particular, the
time integral reduces to sums over bosonic and fermionic Matsubara frequencies. These sums are
evaluated by integrating over appropriate contours using the Orsay representation of the propagators.

\subsection{Schwinger-Dyson Equations}

Our goal is to study the in-medium properties of chiral symmetry breaking in
the models defined in section \ref{models}. Thus nonperturbative methods are required and we
employ the Schwinger-Dyson equations as our organising principle.

\begin{figure}[ht]
\includegraphics[width=6.5cm]{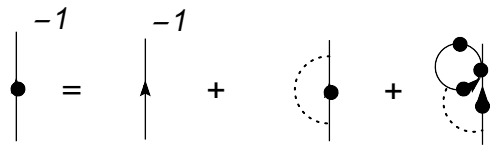}
\caption{Schwinger-Dyson Equation for the full fermion propagator in potential models. Minus signs are not made explicit.}
\label{SDE1}
\end{figure}

\begin{figure}[ht]
\includegraphics[width=8.5cm]{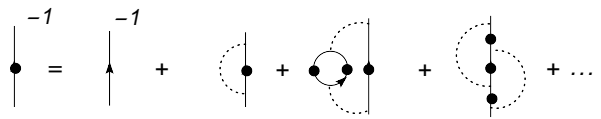}
\caption{Expanded Schwinger-Dyson Equation for the full fermion propagator.}
\label{SDE2}
\end{figure}

The Schwinger-Dyson equation for the full fermion propagator in potential models such as those
employed here is shown in Fig. \ref{SDE1}. The dashed line in this figure represents an 
application of the instantaneous interaction. Notice that this line is not dressed and does not
have an analogous Schwinger-Dyson equation because it is not dynamical.  Inserting the tree level
approximation to the fermion four-point vertex of Fig. \ref{SDE1} yields the equation represented in Fig. \ref{SDE2}. This equation represents a series of diagrams that yield the rainbow-ladder approximation to the fermion propagator, corrections to the rainbow-ladder approximation generated by a dressed 
vacuum polarisation insertion, and corrections due to vertex dressing.

\begin{figure}[ht]
\includegraphics[width=3.5cm]{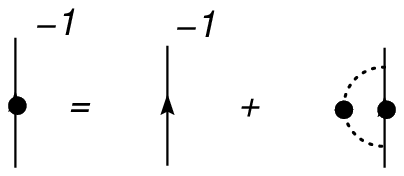} \qquad \qquad
\includegraphics[width=3.5cm]{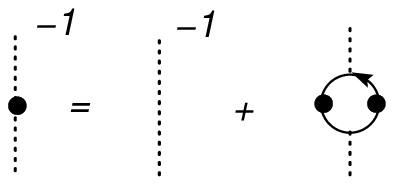}
\caption{Gap Equations: summing polarisation insertions in the truncated Schwinger-Dyson equations.}
\label{SDE3}
\end{figure}

If one neglects vertex correction diagrams, it is possible to sum all dressed vacuum polarisation insertions by rewriting the equation represented in Fig. \ref{SDE2} as shown in Fig. \ref{SDE3}. These coupled equations, called the gap equations,  form the
starting point for our investigation of dynamical mass generation. It is important that the vacuum polarisation fermion loop of the second equation utilises dressed fermion propagators. Failure to do so would yield a branch cut that signifies the decay of dressed fermions to bare fermions -- which is clearly physically unreasonable.

The second gap equation (Fig. \ref{SDE3}) implements the ``ring approximation'' to the full interaction potential. This approximation was first employed by Brueckner and Gell-Mann\cite{ring} to solve the infrared divergence problem in the electron self energy of the degenerate electron gas. As we shall discuss shortly, it will serve the same purpose here. We define the polarisation as

\begin{equation}
\Pi(k_0, k) = \frac{1}{2\beta} n_f \sum_n \int \frac{d^3 p}{(2\pi)^3} 
{\rm tr}[\gamma_0 S(k) \gamma_0 S(p+k)]
\label{PiDef}
\end{equation}
where $n_f$ is the number of quarks (considered of equal mass in the following), $S$ is the 
full quark propagator (discussed more fully below), and $k$ is a four vector defined as

\begin{equation}
k^\mu = (i\omega_n + \mu, \vec k),
\end{equation}
the fermionic Matsubara frequency is given by $\omega_n = (2 n + 1)\pi/\beta$, and $\mu$ is the quark chemical potential. The colour trace yields the factor of $\frac{1}{2}$ in Eq. \ref{PiDef}. The expression for the polarisation is usually split into vacuum and matter components,with the vacuum contribution being defined as $\Pi(k;T\to 0, \mu \to 0)$. Renormalisation follows the standard vacuum formalism and hence only affects $\Pi_{\rm vac}$.

With this definition the ring potential is given by

\begin{equation}
V_{\rm ring}(q_0,\vec q) = \frac{V(\vec q)}{1 - \Pi(q_0,\vec q) V(\vec q)}.
\end{equation}

To establish contact with well-known results we note that 

\begin{equation}
\lim_{p\to 0}\Pi(p_0=0,p) \equiv -m_g^2\, n_f = -\left(\frac{T^2}{6} + \frac{\mu^2}{2\pi^2}\right) n_f
\end{equation}
in the case of a massless bare quark. Of course the full computation must be made with dressed quark propagators and hence forms part of the coupled gap equations.

Incorporating polarisation effects in the gap equations can be very important. For example, quantum electrodynamics in three dimensions is a (logarithmically) confining theory; however, including polarisation effects due to massless fermions {\it completely screens} the confinement potential, leaving a Coulombic heavy fermion interaction\cite{QED3}.
Similarly, increasing the number of quarks in QCD eventually drives the theory into a conformal window with no confinement\cite{BZ}.

\subsection{The Gap Equations}

The diagrams of Fig. \ref{SDE3} represent four coupled integral equations; three involve the 
scalar functions defined in the general expression for the in-medium inverse quark propagator:

\begin{equation}
S^{-1}(k) = i(\omega_n - i \tilde\mu) \gamma_0 - \vec \gamma \cdot \vec k A - B.
\label{Sdefn}
\end{equation}
The scalars $\tilde\mu$, $A$, and $B$ are functions of $k_0$ and $|\vec k|$. Note that $V_{\rm ring}$ depends on $k_0$ and $\vec k$, substantially complicating the solution to the gap equations. However, the dominant infrared contribution to the ring potential is obtained in the static
limit, $k_0 \to 0$ and we employ this limit in the following.  Under these conditions the gap
equations become:

\begin{eqnarray}
A(\vec p) &=& \! 1 + \frac{C_F}{2} \!\! \int\frac{d^3q}{(2\pi)^3} V_{\rm ring}(\vec p-\vec q) \frac{A_q}{E_q} \frac{\vec p \cdot \vec q}{p^2} \Theta(q)\nonumber \\
B(\vec p) &=& \! m + \frac{C_F}{2} \!\! \int\frac{d^3q}{(2\pi)^3} V_{\rm ring}(\vec p-\vec q) \, \frac{B_q}{E_q} \, \Theta(q) \nonumber \\
\tilde \mu(\vec p) &=& \mu + \frac{C_F}{2} \!\! \int\frac{d^3q}{(2\pi)^3} V_{\rm ring}(\vec p-\vec q) [n(q) - \bar n(q)] \nonumber \\
E_p^2 &=& \! A_p^2 \, p^2 + B_p^2.
\label{GapEqs}
\end{eqnarray}
We have introduced the colour factor $C_F = (N^2-1)/(2N)$.
The thermodynamic function is defined in terms of the quark densities as

\be
\Theta(q) = 1 - n(q) - \bar n(q)
\ee
with
\begin{equation}
n(p) = \frac{1}{\exp(\beta (E_p-\tilde\mu)) + 1}
\end{equation}
and
\begin{equation}
\bar n(p) = \frac{1}{\exp(\beta (E_p+\tilde\mu)) + 1}.
\end{equation}

Similar equations have been considered by Koci{\'c} \cite{kocic} and more recently in
Refs. \cite{GS, leo}. Koci{\'c} presents analytic results for a contact model very similar to ours, but without consideration of vacuum polarisation effects. All three papers consider the linear confinement model at zero temperature; Ref. \cite{GS} incorporates ring
corrections and examines the effect on charmonium dissociation, but neglects the 
vacuum part of
the polarisation without comment. A contact model is considered at finite temperature and density in 
Ref. \cite{david}; additional related studies are listed in Ref. \cite{other}.

In the case of the contact potential or the ring contact potential in the static and long wavelength limits (discussed below) these equations simplify further to 

\begin{eqnarray}
A &=& 1, \nonumber \\
B &=& m + \frac{C_F}{2} \int\frac{d^3q}{(2\pi)^3}V^{\rm con}_{\rm ring}(\vec 0) \frac{B(q)}{E_0(q)} \Theta_0(q), \nonumber \\
\tilde\mu &=& \mu + \frac{C_F}{2} \!\! \int\frac{d^3q}{(2\pi)^3} V^{\rm con}_{\rm ring}(\vec 0) [n(q) - \bar n(q)] \nonumber \\
E_0^2(p) &=& p^2 + B^2, \\
\label{GapContactEqns}
\end{eqnarray}
where $\Theta_0$ is defined in terms of $E_0$.

More generally, the ring contact case is as complicated as the linear model since the interaction becomes

\begin{equation}
V^{\rm con}_{\rm ring}(\vec p) = \frac{1}{\Lambda^2} \frac{\lambda}{1 - \lambda \frac{\Pi(0,\vec p)}{\Lambda^2} }.
\end{equation}

\noindent
Since the chief role of the polarisation function is to regulate the infrared divergence 
that appears in the gap equations (discussed in the next section), it is appropriate to consider the approximation:

\begin{equation}
V_{\rm ring}(\vec q) = \frac{V(\vec q)}{1 - \Pi(0,\vec q \to 0) V(\vec q)}.
\label{VringApprox}
\end{equation}
Indeed, as Eq. \ref{VringApprox} illustrates, in the linear case $V(\vec q)$ dominates $V_{\rm ring}$ for $q^2 \gg m_g \sqrt{b}$ and it is reasonable to use the zero momentum limit of the polarisation function in the gap equations. We shall confirm this in the following. In this limit the vacuum contribution to the polarisation function vanishes and we need only consider the matter contribution. In fact no ultraviolet divergences remain in the linear model and we do not consider renormalisation. Of course the contact model must be cutoff at the scale $\Lambda$, as previously indicated.

In the contact case, this approximation implies that $V_{\rm ring}$ is a
constant and hence $B$ is a constant. Thus the integral equation for $B$ in 
the contact case becomes a relatively simple algebraic equation, greatly simplifying the problem.

In the low momentum limit the explicit expression for $\Pi$ becomes

\be
\Pi(0,\vec q \to 0) = -4 n_f \beta \int \frac{d^3 k}{(2\pi)^3} [2+{\rm e}^{\beta E(k)} + {\rm e}^{-\beta E(k)}]^{-1}
\ee
where $E(k)$ depends on $A$ and $B$ and is naively infrared divergent (cf. Eqs. \ref{GapEqs} and section \ref{sectD}).

Finally, the dynamical quark mass is given by

\be
M(p) = \frac{B(p)}{A(p)}
\ee
and a gap equation can be derived for this function directly:

\begin{eqnarray}
M(p) &=& m + \frac{C_F}{2} \int \!\! \frac{d^3 q}{(2\pi)^3} V_{\rm ring}(\vec p - \vec q) \cdot \nonumber \\
&&  \left[ \frac{M(q)}{E_0(q)} - \frac{M(p)}{E_0(q)} \frac{\vec p \cdot \vec q}{p^2}\right] \Theta(q)
\label{MEq}
\end{eqnarray}
where now $E_0(q) = \sqrt{q^2 + M(q)^2}$. It is important, however, to note that $\Theta$ still depends on the full single particle energy, $E(q)$.

\subsection{Infrared Behaviour of the Gap Equations}
\label{sectD}

In the absence of polarisation effects 
the linear model gap equations of Eq. \ref{GapEqs} are naively infrared divergent. Specifically as 
$\vec q \to \vec p + \vec \delta$ on the right hand side, the potential diverges as $V(\delta) \sim \delta^{-4}$. Thus the quark energy $E(q)$ is divergent, the thermodynamic function approaches  
unity, 
and temperature effects disappear from the in-medium gap equations. 
We stress that it is not possible to write the equations in terms of IR-finite quantities,
in contrast to the zero temperature case.
This nonsensical result was first noted by Davis and Matheson\cite{DM} who suggested that one should make the {\it ad hoc} replacement 

\be
E(q) \to E(q) - E(0)
\ee
in the expression for $\Theta(q)$. This removes the infrared divergence and yields a sensible in-medium gap equation.

An alternative suggestion was made by Alkofer {\it et al.}\cite{AAL}, who simply replaced $E(q)$ with $E_0(q)$.
The utility of this Ansatz is that the equation for $M$ is infrared finite as evidenced in Eq. \ref{MEq}. We shall denote this procedure as `AAL' in the following.

Finally, the Orsay group has argued that infinite quark energies are physically reasonable and reflect the absence of individual quarks in the physical spectrum\cite{orsay}. They therefore reformulate the thermodynamic trace to sum over colour singlet states only. This greatly complicates the thermodynamic trace but eventually yields a simple result: momentum space integrals such as
appearing in Eqs. \ref{GapEqs} are restricted to $q \neq p$, thereby eliminating the infrared divergence.

None of these resolutions seem appropriate.  The substitutions of Refs. \cite{DM} and \cite{AAL} are completely {\it ad hoc}. However, it is possible that they are reasonable approximations, and we investigate this in the next section. The Orsay approach is physically motivated, however, it should not be necessary to explicitly remove nonsinglet degrees of freedom from the thermodynamic trace because they are automatically removed by the Boltzmann factor.

We note the following: (i) infrared divergences are normally removed by considering additional physical effects (such as bremsstrahlung) or by summing additional classes of diagrams (as with the resolution of the IR divergence problem in the degenerate electron gas\cite{ring}  or in the 
hard thermal loop (HTL) partial resolution of gauge-variance and other problems in thermal 
QCD\cite{BP}) (ii) `artificial' infrared divergences sometimes appear in problems involving confinement potentials\cite{ex}.

We examine  both issues in the following. Specifically, issue (i) has been addressed by 
summing the ring diagrams in the gap equations. This has the effect of removing the infrared
divergence in all expressions, as evident in Eq. \ref{VringApprox}. Surprisingly, issue (ii) is also relevant here. Indeed, the gap equations are trivially IR regulated by subtracting zero. For example, the third of Eqs. \ref{GapEqs} can be rewritten as 

\be
B_p =  m + \frac{C_F}{2} \!\! \int\frac{d^3q}{(2\pi)^3} V_{\rm ring}(\vec p-\vec q) \left[ \frac{B_q}{E_q} \Theta(q) - \frac{B_p}{E_p} \Theta(p) \right].
\label{BRegEq}
\ee
While this yields a formally IR-finite equation, we find that it is still strongly dependent on
the infrared regime, thus it remains useful to employ the ring potential in the formalism when 
performing numerical computations.

We examine the regulated gap equations with the bare and ring potentials in the following section.

\section{Phase Structure}

Our results will be presented as plots of the dynamical mass at zero momentum as a function of chemical potential and temperature. Thus we determine the phase structure of the contact and confinement models. As far as we know these are the first computations of these phase diagrams for confining potentials.
The results for all potentials with ring corrections are also new.

The coupled gap equations are solved with a variety of techniques. Our typical approach is to
compute on a momentum grid. Unknown functions $A$, $B$, and $V_{\rm ring}$ are determined by minimizing an appropriate functional that represents the gap equations using a modified Levenburg-Marquardt algorithm.
Alternatively, we attempt iterative solutions (which are often not stable) and combinations of iteration and minimisation.
We also employ a variety of analytic approximations to quantities such as $M$ and $V_{\rm ring}$ and
confirm the more complete results. For example, computing the full momentum dependence of $\Pi$ is time-consuming
and it is useful to expand it as a Taylor series in momentum. The results presented here represent several months
of single core computer time.

\subsection{Contact Model Results}

The contact model is defined in terms of the coupling $\lambda$ and ultraviolet cutoff $\Lambda$. It is IR-finite and hence the only issues here are the effect of the ring interaction and, of course, the new phase structure shown below. Contact models such as ours have a critical coupling $\lambda_c$, below which no chiral symmetry breaking occurs. Since this is a zero-temperature property of the model and $\Pi_{\rm mat}$ vanishes in this limit, the critical
coupling is the same for the bare and ring models:

\be
\lambda_c = 6 \pi^2.
\ee

To set conventions we choose to work in the broken phase, setting $\lambda = 1.5\lambda_c$ and then fixing the scale $\Lambda = 370$ MeV to yield reasonable approximations for the dynamical 
quark mass
and chiral restoration temperature, $M(T=0,\mu=0) \approx 260$ MeV and $T_c \approx 150$ MeV. The chiral condensate is given by the expression

\be
\langle \bar \psi \psi \rangle = -\frac{3}{\pi^2}\int k^2 dk \, \frac{B(k)}{E(k)}
\ee
and is numerically (-150 MeV)$^3$, which is approximately a factor of two too small (in linear dimension). 
Finally the critical chemical potential is approximately $300$ MeV.  The numerical values of these observables shift slightly
with the coupling, for example for $\lambda = 3 \lambda_c$ one finds $\Lambda = 140$ MeV, $T_c = 135$ MeV, 
and $\mu_c = 250$ MeV.

It is the fact that the zero temperature gap equation has a critical coupling that drives the finite temperature and density phase transitions. Specifically, the presence of the thermodynamic function $\Theta$ at $T>0$ can be modelled as an effective coupling:

\be
\lambda(T,\mu) = \lambda \, \langle \Theta(q;T, \mu) \rangle
\ee

\noindent
where the angle brackets denote some average over the integrand of the gap equation. Once $\lambda(T,\mu)$ drops below $\lambda_c$ the model makes the transition to the symmetric phase.

\begin{figure}[ht]
\includegraphics[width=7cm,angle=-90]{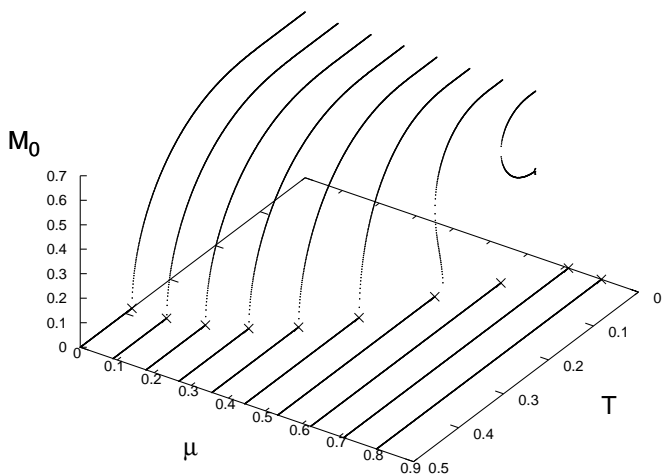}
\caption{Dynamical mass vs. temperature and density for the contact model ($\lambda = 1.5\lambda_c$, all quantities in units of $\Lambda$).}
\label{M-T-mu-contact-1.5}
\end{figure}

The phase diagram for the bare contact model is shown in Fig. \ref{M-T-mu-contact-1.5}. In the following we
will refer to $M_0$ defined as $M(k\to 0;T,\mu)$. This represents a `constituent quark mass' and is a useful and simple proxy for the chiral condensate. We note that the simplicity of the contact gap equations permit a detailed analysis of the phase structure, including a precise determination of the order of the phase transition. This is useful since, as the figure illustrates, the phase structure is quite complex. We find that the 
chiral symmetry restoration phase transition is second order for all chemical potentials below  a critical density

\be
\mu_\star(\lambda=1.5\lambda_c) \approx 0.53 \Lambda.
\ee
For higher chemical potential the transition is first order. Furthermore, a second solution
to the gap equations develops
 -- these are the lower lines seen in the figure between $\mu_\star$ and $\mu= 0.9$. These solutions indicate the presence of a first order phase transition; however, they are unphysical in that they have higher free energy, as demonstrated by their lack of continuity with the low-$\mu$ solutions.

\begin{figure}[ht]
\includegraphics[width=7cm,angle=-90]{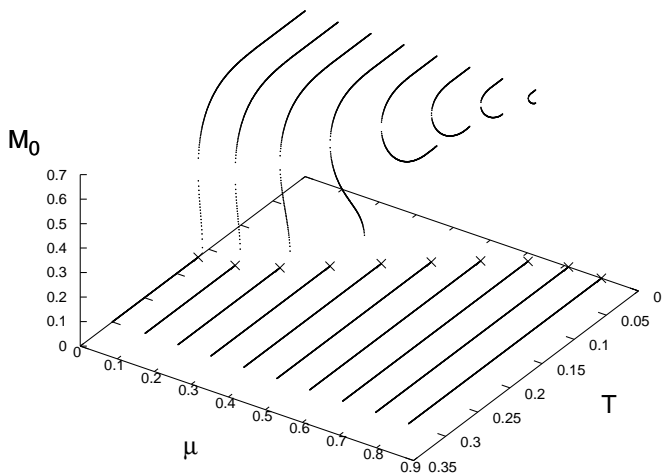}
\caption{Dynamical mass vs. temperature and density for the contact model with the static, zero momentum ring potential ($\lambda = 1.5 \lambda_c$, all quantities in units of $\Lambda$).}
\label{M-T-mu-contact-ring}
\end{figure}

Including the zero-momentum vacuum polarisation function (with $n_f = 1$) in the gap equations induces nontrivial dynamical mass, temperature, and chemical potential dependence in the kernel of the gap equation. The results are presented in Fig. \ref{M-T-mu-contact-ring}. It is apparent that the additional mass dependence causes the bifurcated solution to exist for all chemical potential. The phase transition is first order for all 
values of temperature and chemical potential. Furthermore, the numerical value of the critical temperature is strongly affected: 

\be
T_c(\mu=0; \lambda=1.5\lambda_c; {\rm bare}) = 0.38 \Lambda
\ee
drops to

\be
T_c(\mu = 0; \lambda = 1.5 \lambda_c; {\rm ring}) = 0.17 \Lambda.
\ee
For $\lambda = 3 \lambda_c$ the analogous results are $T_c({\rm bare}) = 0.95\Lambda$ and $T_c({\rm ring}) = 0.30\Lambda$.
Of course, $\mu_c$ does not change because it is defined at $T=0$ and $\Pi=0$ at this point. Thus the shape of the critical region and the phase transition are strongly affected by moving beyond the bare interaction approximation in the gap equation.

\begin{figure}[ht]
\includegraphics[width=7cm,angle=-90]{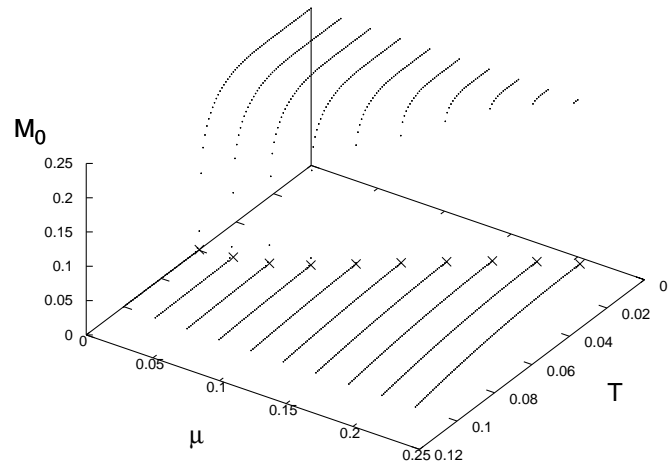}
\caption{Dynamical mass vs. temperature and density for the contact model with the static ring potential ($\Pi_{\rm mat} + \Pi_{\rm vac}$, $\lambda = 1.5 \lambda_c$, all quantities in units of $\Lambda$).}
\label{M-T-mu-contact-ring-full}
\end{figure}

Finally, we consider the case where the full momentum dependence is retained in the vacuum polarisation, leading to 
a momentum dependent ring potential. The results are shown in Fig. \ref{M-T-mu-contact-ring-full}. We see
that the dynamical fermion mass has dropped to $M = 0.23 \Lambda$, as have the critical temperature and 
chemical potential $T_c = 0.06 \Lambda$, and $\mu_c = 0.22\Lambda$. However, upon rescaling $\Lambda$ we find that all of these quantities
remain essentially unchanged. Furthermore, it appears that the phase transition is first order throughout the
phase diagram, as it  is in the zero-momentum ring case (determining the order of the phase transition is
difficult with momentum dependent interactions because algebraic methods are not applicable). Thus it
appears that including the full momentum dependence of the polarisation function and ring potential alters
the numerical values of quantities in units of the cutoff, but leaves physical quantities approximately 
invariant.

We have also studied the $\lambda$-dependence of the phase diagram for the `bare' contact model.  We find that all quantities 
scale roughly linearly with $\lambda$. For example, to a good approximation

\be
M_0(\lambda) = 0.011 \cdot \Lambda \lambda \cdot \left(1 - {\rm e}^{-0.028(\lambda-\lambda_c)}\right).
\ee
Furthermore, the relative strengths of the dimensional quantities seem to be roughly independent of the coupling, and we find:

\be
\mu_c(\lambda) \approx  M_0(\lambda) \approx 2 T_c(\lambda) \approx 2 \mu_\star(\lambda)
\ee

\begin{figure}[ht]
\includegraphics[width=6cm,angle=-90]{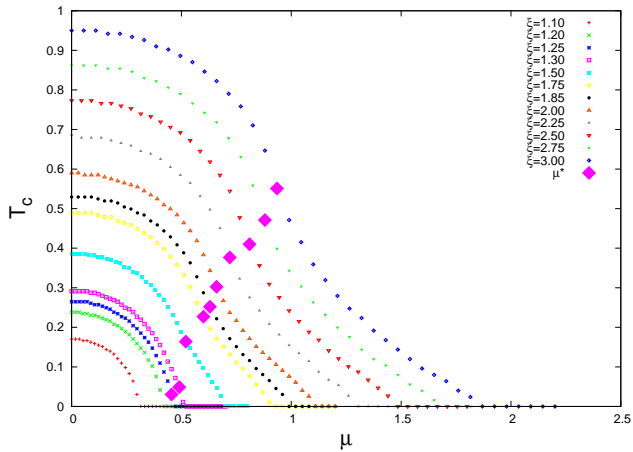}
\caption{The bare contact interaction phase diagram for a variety of couplings ($\xi \equiv \lambda/\lambda_c$).}
\label{phase-diagram-contact}
\end{figure}

The coupling dependence of the phase diagram for the bare interaction case is illustrated in 
Fig. \ref{phase-diagram-contact}. The approximate linear scaling is evident. This is also true
for the critical point $\mu_\star$ down to $\lambda \approx 1.2 \lambda_c$. Below this point
the first order phase transition disappears and the entire diagram represents a second order phase transition.

\subsection{Confinement Model Results}

The confinement models exhibit the full IR singularities that have driven the discussion. Unfortunately, the models are much more difficult to analyse and one must rely on accurate numerics to establish the order of 
phase transitions.  The relation of the chiral restoration and deconfinement transitions is also of interest. Longstanding lattice gauge theory results indicate that these transition temperatures coincide\cite{latt}. Since the confinement potential is fixed bare models cannot reproduce 
this behaviour. Indeed
Davis and Matheson have argued\cite{DM} that all confining potentials break chiral symmetry and since the only in-medium effect is the addition of the factor $\Theta$ in the gap equation kernels, it is feasible that there can be {\it no} transition to the chirally symmetric phase. Nevertheless, Alkofer {\it et al.} have numerically found a phase transition.

\begin{figure}[ht]
\includegraphics[width=7cm,angle=-90]{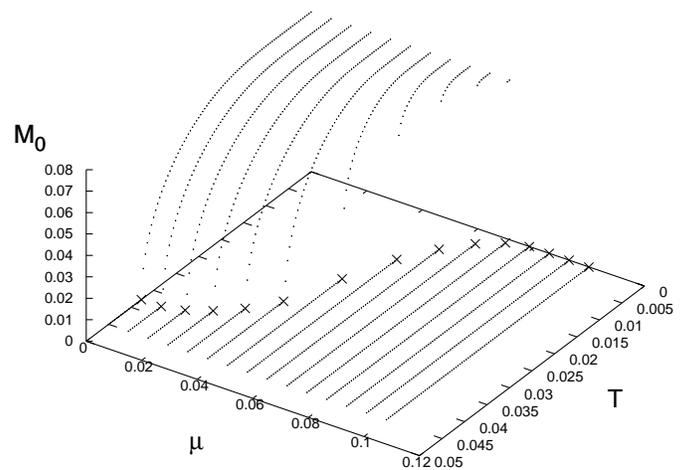}
\caption{Dynamical mass vs. temperature and density for the bare linear AAL confinement model.}
\label{M-T-mu-V-AAL}
\end{figure}

As shown in Fig. \ref{M-T-mu-V-AAL} we confirm the existence of a phase transition in the confinement model with the AAL prescription (and we extend their results to the $T-\mu$ plane). We speculate that
the argument of Davis and Matheson has failed because for high temperature and low momentum the thermodynamic
function is approximately $\Theta(q,T,\mu=0) \sim q/T$, and hence the effective confining potential is
replaced by $V \sim q^{-4} \to q^{-3}$, which is not confining. Perhaps this is sufficient to drive the observed behaviour.

Numerical values for the linear potential are $T_c \approx 38$ MeV, $\mu_c \approx 75$ MeV, $M_0 \approx 80$ MeV, and the condensate is determined to be $(-\langle\bar \psi \psi\rangle)^{1/3} \approx 110$ MeV. Note that we interpret the dynamical mass as a constituent quark mass. All of these quantities fall below their phenomenologically expected values; however, using a string tension of about 1.8 GeV$^2$ brings them all into reasonable agreement. In fact, 
Alkofer {\it et al.} employed such a large string tension in their computation. This is unfortunately at odds
with well established hadron phenomenology; a point which we will discuss further below.

\begin{figure}[ht]
\includegraphics[width=6cm,angle=-90]{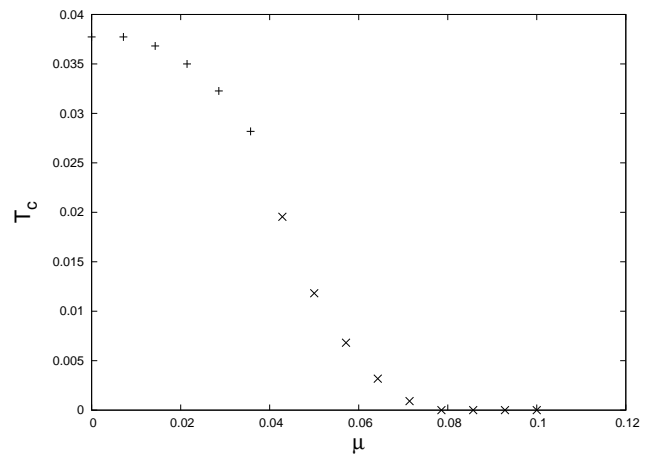}
\caption{The bare linear AAL phase diagram.}
\label{phase-diagram-linear}
\end{figure}

The chiral restoration transition line is plotted in Fig. \ref{phase-diagram-linear}. A clear inflection
point is visible at $(T_\star,\mu_\star) \approx (20,43)$ MeV. We identify the right side of this point as a region
of first order phase transitions, while the left side is second order. This identification is based on
the continuity of $M$ displayed on Fig. \ref{M-T-mu-V-AAL}, the evident first order phase transition seen at $T=0$, the relative lack of stability of the solution algorithm for large $\mu$, and the appearance of secondary solutions (not shown) for $\mu > \mu_\star$. The existence of
a tricritical point is in keeping with expectations for QCD\cite{BBB}.

The results of Fig. \ref{M-T-mu-V-AAL} have been computed with the {\it ad hoc} AAL
prescription. We now consider modifying the infrared behaviour of the quark interaction with the 
vacuum polarisation diagram of Fig. \ref{SDE3}. Our first study is in the static, low momentum limit with $n_f=1$. In this 
case the vacuum contribution to $\Pi$ vanishes and we need not consider renormalisation of the model.

\begin{figure}[ht]
\includegraphics[width=7cm,angle=-90]{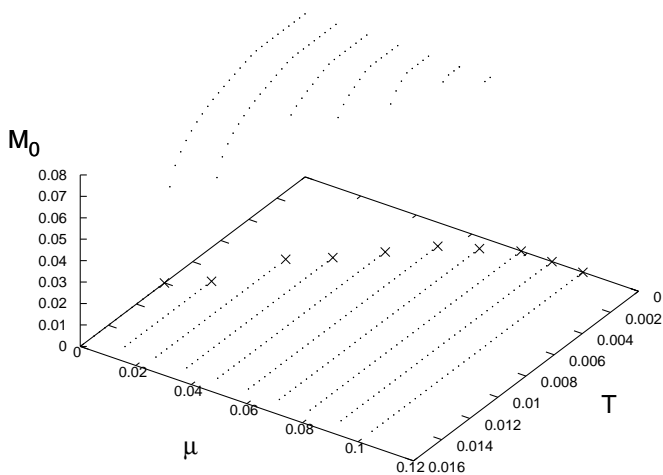}
\caption{Dynamical mass vs. temperature and density for the linear static long wavelength ring approximation}
\label{linear_ring_q=0}
\end{figure}

Our results are shown in Fig. \ref{linear_ring_q=0}. The AAL and ring results must agree at the origin.
The surfaces at $\mu=0$ look very similar, however the critical temperature has dropped from
38 MeV to 10 MeV. The critical chemical potential remains fixed at $\mu_c \approx 75$ MeV. Thus the shape of
the phase diagram has been severely distorted by incorporating vacuum polarisation in the model. Furthermore, as with
the contact model, it appears that all phase transitions are now first order, and the tricritical point no longer
exists. Thus, as anticipated, including dynamical quarks in the model can have a dramatic effect on thermodynamic
properties.

The vacuum polarisation function introduces explicit temperature and density
dependence to the quark interaction, which raises the possibility of explicit quark deconfinement in the model.
It is possible that this dependence causes the potential to deconfine at a critical temperature. However, it is more likely that the potential deconfines for all nonzero temperature. Indeed, 
in perturbation theory one can approximate the ring potential as

\be
V_{\rm ring}(q,T,\mu=0) \approx \frac{6\pi b}{q^4 + \pi b T^2}.
\ee
The Fourier transform of this potential is linear when $T=0$. When $T>0$ the potential is
linear at small distances, has a transition region at $r \sim (\pi b T^2)^{-1/4}$, and approaches zero at large distances, so that deconfinement is natural, although not sudden.
A careful analysis of deconfinement awaits a study of QCD. 

Finally, we consider incorporating the full momentum dependence of the polarisation function in the gap
equations. 
It is important to note that the 
vacuum contribution to the polarisation function is not zero for non-zero momentum. This contribution must
be renormalised at the expense of introducing another parameter to the model. In practice, the relatively good
agreement with heavy quark spectroscopy would be ruined by this procedure. We therefore consider $\Pi_{\rm vac}$
to already be incorporated in the model interaction, and simply consider matter contributions to the ring
potential. 

Unfortunately, the numerical solution of the 
gap equations is substantially complicated in this case. Solving the full set of coupled integral equations can
be time consuming and unstable. We have found that setting $A(q)$ to be large provides a very good approximation to the full gap equations, and this approximation has been employed for most of the linear ring results presented here. An additional complication is provided by the requirement to
compute the polarisation
at many grid points at each step of the computation. However, we have found that expanding $\Pi(q)$ to ${\cal O}(q^4)$
is  very accurate, and this speeds the solution tremendously. 

In the end, the results are very
similar
to those of the zero momentum case, as anticipated above. For example, the dynamical mass $M(k=0, T=0, \mu=0)$ is lowered by approximately 1 MeV when incorporating the momentum dependence of the vacuum polarisation. Similarly, the critical temperature remains the same to the accuracy we compute.

The results presented in this section are for the linear potential of Eq. \ref{LinEq}. We have performed the same computations with the Richardson confinement model of Eq. \ref{RichEq}. This permits examining the dependence of our results on the details of the confinement model. In particular,
the Richardson potential incorporates a running Coulomb interaction, and it has been speculated that this can enhance the chiral condensate. We find, however, that the Richardson potential yields
nearly identical results as the linear potential. For example, the AAL Richardson dynamical mass
is only 1 MeV less than the corresponding linear mass. Similarly, the critical temperature drops 0.7 MeV and the critical density raises 3.2 MeV.

\section{Conclusions}

We have examined bare and ring versions of a contact model that mimics the structure of QCD in Coulomb gauge. First results for the phase diagram of the ring contact model are presented here. The bare model exhibits interesting behaviour, including approximate linear scaling
of $T_c$, $\mu_c$, and the dynamical mass with the coupling. The numerical values of these quantities can be placed in rough agreement with QCD expectations with an appropriate choice of the cutoff scale. Surprisingly, we find a tricritical point for all models with $\lambda > 1.2 \lambda_c$, again in keeping with expectations for QCD. However, incorporating ring-type Schwinger-Dyson equations in the
formalism changes these conclusions dramatically. Ring values for $\mu_c$, $M_0$ and the condensate do not change substantially, however, the critical temperature drops by a factor of three,
 ruining the
phenomenology. Furthermore, the region of second order phase transitions disappears and only a first order phase transition appears in the phase diagram.  Given the strong $n_f$-dependence expected
in gauge theories, 
perhaps the strong effects seen here should not be surprising.

For the bare linear and Richardson models with the AAL infrared prescription we confirm the existence of a
second order phase transition  at small chemical potential. For chemical potential larger than
$\mu_\star \approx 43$ MeV the phase transition becomes first order.
The appearance of {\it any} phase transition is somewhat surprising, since it is 
in conflict with the reasonable expectations of Davis and Matheson. The numerical values for 
the dynamical mass, chiral restoration temperature and density, and chiral condensate are  all in 
agreement with QCD expectations if the string tension is increased to a 
 value of 1.8 GeV$^2$. Unfortunately, this is in severe conflict with well-established quark model
phenomenology and lattice gauge results that require a string tension of approximately 0.2 GeV$^2$.
It is thus apparent that the simplest confinement models cannot both reproduce
thermodynamic and spectroscopic quantities with any reliability.
Of course this conclusion depends on the approximations we have made. However, the large discrepancy
seems difficult to overcome and we expect that simple confinement models are incapable of describing
in-medium properties of QCD. These conclusions hold for both linear and Richardson models.

We have noted that two resolutions to the confinement model infrared problem exist: a direct
physical resolution is to employ the ring approximation and a simple mathematical resolution
involves the proper regularisation of the Fourier transform of the linear potential. We find
sufficient numerical infrared instability to warrant employing the physically reasonable, and
presumably more accurate, ring Schwinger-Dyson equations. Results in the static/low momentum and
static/finite momentum limits were found to coincide very well. As with the contact interaction, ring diagrams induce
strong effects in thermodynamic observables, causing the critical temperature to drop by a factor of four
and changing all phase transitions to first order.


Pisarski and McLerran have recently used the large $N_c$ limit to argue that a `quarkyonic'
QCD phase may exist\cite{pml}. This phase is proposed to exist at high chemical potential and low
temperature and to be characterised by chirally symmetric but confined quark matter.  
(We note that the argument is supported by observations that chiral symmetry breaking should be 
irrelevant
high in the zero temperature hadron spectrum\cite{chiral}.)
The large $N_c$ limit suppresses quark loop effects, thereby yielding an interaction that
is independent of the chemical potential. Thus the linear no-ring model presented here can be 
considered an implementation of the large $N_c$ scenario. Our results then confirm the idea that
a confining but chirally symmetric phase can exist. Of course, decreasing the number of colours 
implies an increasing importance of vacuum polarisation with commensurate changes in phenomenology,
as noted above. A full assessment of the quarkyonic matter scenario must await a nonperturbative
investigation of more realistic models of QCD.

The application of finite temperature and density Schwinger-Dyson methods to the
QCD Hamiltonian will be of great interest. As illustrated here, one of the overriding considerations will be the construction of a robust truncation scheme for Hamiltonians with infrared enhanced interactions. Indeed, the Ward identities relate vertices and propagators, so one expects that
if corrections to propagators are necessary, one should truncate full vertices very carefully. Of course this observation is borne out by the experience with the effective field theory generating the HTL limit of QCD.

\acknowledgments
The authors are grateful to Dan Boyanovsky for many discussions on thermal field
theory.
This research was supported by the U.S. Department of Energy under contract
DE-FG02-00ER41135.


\begin{thebibliography}{99}
\bibitem{ex}
A simple example is the following: the integral $\int d^3r \, r \exp(-r^2/2)$ is well-defined; however its momentum space equivalent is $\int d^3 k / (2\pi)^3 (-8 \pi/k^4) (2 \pi)^{3/2} \exp(-k^2/2)$ which is infrared divergent. The problem, of course, can be traced to the naive expression for the Fourier transform of a linear potential.
\end{thebibliography}
\end{document}